\documentclass[runningheads]{llncs}

\usepackage[T1]{fontenc}
\usepackage{graphicx}
\usepackage{bm}
\usepackage{amsfonts}
\usepackage{amsmath}
\usepackage[inline]{enumitem}
\usepackage{xspace}
\usepackage{dsfont}
\usepackage{xcolor}
\usepackage{multirow}
\usepackage{subfigure}
\usepackage[pagebackref=false,breaklinks,colorlinks]{hyperref}
\usepackage{color,colortbl}

\usepackage{cleveref}

\newcommand{\querywise}{{QueryWise}\xspace}
\newcommand{\qw}{{QW}\xspace}
\DeclareMathOperator*{\argmax}{argmax}

\newcommand{\myfirstpara}[1]{\par \noindent \textbf{#1.}}
\newcommand{\mypara}[1]{\vspace{0.2em} \myfirstpara{#1}}

\def\L{\mathcal{L}\xspace}

\begin{document}
\title{Assessing Risk of Stealing Proprietary Models for Medical Imaging Tasks}

\author{Ankita Raj\inst{1}\and
	Harsh Swaika\inst{1}\and
	Deepankar Varma\inst{2}\thanks{Work done as an intern at Indian Institute of Technology Delhi}\and
	Chetan Arora\inst{1}\thanks{Corresponding author}}

\authorrunning{A. Raj et al.}

\institute{Indian Institute of Technology Delhi, India \\
 \and
	Thapar Institute of Engineering and Technology, Patiala, India}

\maketitle   
\begin{abstract}
	The success of deep learning in medical imaging applications has led several companies to deploy proprietary models in diagnostic workflows, offering monetized services. Even though model weights are hidden to protect the intellectual property of the service provider, these models are exposed to model stealing (MS) attacks, where adversaries can clone the model's functionality by querying it with a proxy dataset and training a thief model on the acquired predictions. While extensively studied on general vision tasks, the susceptibility of medical imaging models to MS attacks remains inadequately explored. This paper investigates the vulnerability of black-box medical imaging models to MS attacks under realistic conditions where the adversary lacks access to the victim model's training data and operates with limited query budgets. We demonstrate that adversaries can effectively execute MS attacks by using publicly available datasets. To further enhance MS capabilities with limited query budgets, we propose a two-step model stealing approach termed \texttt{QueryWise}. This method capitalizes on unlabeled data obtained from a proxy distribution to train the thief model without incurring additional queries.  
	Evaluation on two medical imaging models for Gallbladder Cancer and COVID-19 classification substantiate the effectiveness of the proposed attack. 
	The source code is available at \url{https://github.com/rajankita/QueryWise}. 
\end{abstract}

\section{Introduction}

\myfirstpara{Adversarial attacks}
Deep learning models have emerged as a prominent tool in medical imaging, improving diagnosis and accelerating decision-making in clinical tasks. However, these systems pose security and privacy risks.  While threats like adversarial \cite{goodfellow2014explaining} and poisoning attacks \cite{munoz2017towards} have been investigated in medical imaging \cite{xu2021towards,alkhunaizi2022suppressing,pandey2022adversarially}, model stealing attacks remain largely unexplored. 

\mypara{Model stealing attacks}
Model Stealing (MS) attacks, also known as Model Extraction attacks, pose a significant threat to the confidentiality of proprietary machine learning models \cite{kumar2020adversarial,orekondy2019knockoff,pal2020activethief}. In an MS attack, adversaries aim to replicate the functionality of a private machine learning model that operates as a black box and is accessible only through querying, while its weights are hidden from end users. This operational model, known as Machine-learning-as-a-Service (MLaaS), has been adopted by several companies \cite{qure,skinvision}, which provide proprietary machine-learning models to hospitals, healthcare professionals, or end-users for a fee. To safeguard their intellectual property, these companies restrict access to the model's internals, such as the training dataset, model architecture, and hyperparameters, while offering only black-box access to the model. However, in an MS attack, a malicious user tries to replicate the functionality of the black-box model, referred to as the \textit{victim} model, by querying it with a proxy dataset and training a substitute model using the acquired predictions. Given the high labeling costs in medical imaging, often requiring expert knowledge, such attacks could enable competitors to develop their own models at significantly lower costs. This not only jeopardizes the intellectual property of the model owner but also exposes the model to additional threats, such as model inversion attacks.

\begin{figure}[t]
	\centering
	\includegraphics[width=\textwidth]{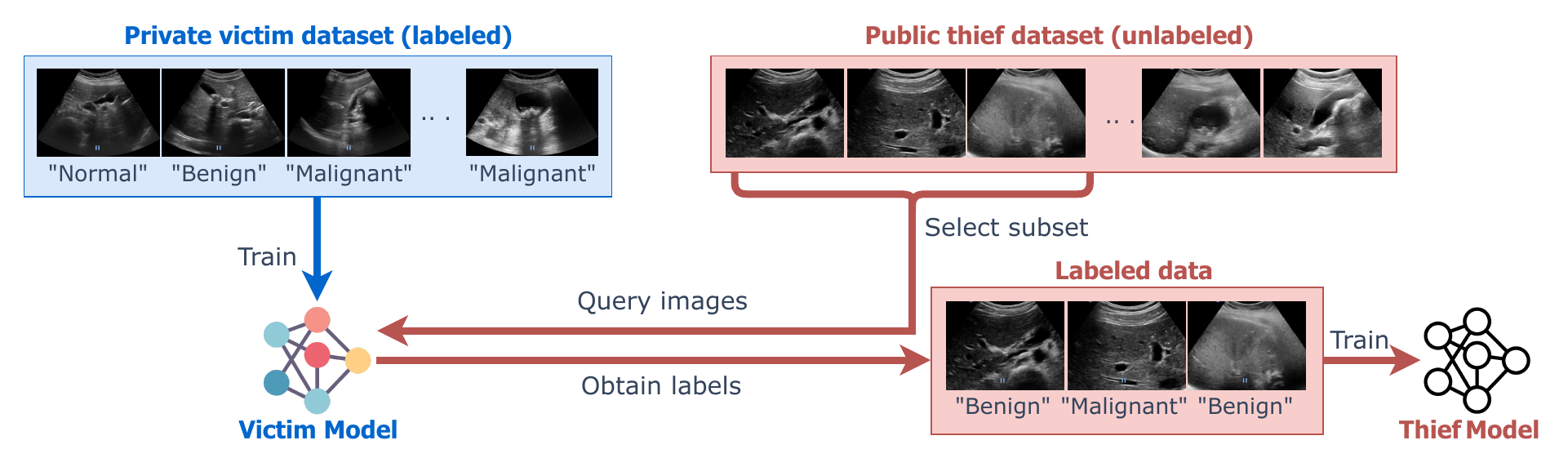}
	\caption{\textbf{Model Stealing Attack setup:} Victim model is a black-box. An adversary samples images from an unlabeled thief distribution and queries labels from the victim model to train a thief model.}
\end{figure}

\mypara{Challenges with stealing medical imaging models}
Despite the significant threat, MS attacks remain under-explored for medical imaging. Existing methods for MS attacks on natural image-based models \cite{orekondy2019knockoff,pal2020activethief,wang2021black,Beetham2023} require thousands or even millions of queries, impractical for medical datasets with limited samples.
While some studies have demonstrated attacks on medical datasets, they are either concerned with stealing text-based models \cite{zhang2020neural} or rely on relaxed assumptions necessitating the acquisition of the full prediction vector from the victim model \cite{orekondy2019knockoff}. Our work involves stealing medical imaging models under a more realistic scenario where a victim model provides only the top-1 prediction, also known as \textit{hard-label}, and an extremely small query budget.
Under these stringent stealing conditions, the limited queried subset fails to adequately cover the victim's input space, resulting in thief models significantly trailing behind in accuracy compared to the victim. 

\mypara{Our proposal}
We note that state-of-the-art (SOTA) MS techniques primarily focus on training the thief model solely on the small queried subset, neglecting a large portion of unused proxy data. We propose \textbf{\querywise}, a novel MS method that leverages both queried/labeled data and the remaining unlabeled data for thief model training. 
In the first step, we train a model exclusively on the labeled data, which we call the \textit{anchor model}, using established MS attack techniques \cite{orekondy2019knockoff,pal2020activethief}. The final thief model is trained in the second step, utilizing both labeled and unlabeled data, with guidance from the anchor model and an additional \textit{teacher model}. The teacher model's weights are updated using an exponential moving average of the thief model's weights to ensure stable predictions \cite{tarvainen2017mean}. 
Since the thief's proxy data lies out-of-distribution (OOD) with respect to victim model's distribution, hard-labels queried from the victim may lack meaningful information and thus have limited utility. A key insight of our work is that by generating soft pseudo-labels for the unlabeled data using a combination of the anchor and teacher models, our approach facilitates the thief model in capturing label correlations with other samples more effectively.
\cite{jagielski2020high} showed the benefits of using unlabeled data for MS attacks, but it remained unclear if these benefits applied when the proxy data differed from the victim data. \cite{wang2021black} employed an iterative training method with random erasure to generate pseudo-labels for unlabeled data. Our method improves upon this by combining pseudo-labels from an anchor model and a weighted-averaged teacher model, retaining only high-confidence labels for superior supervision. Crucially, we use the anchor model to guide the thief model on labeled data (see \Cref{student_training}).

\mypara{Contributions}
We make the following key contributions:
\begin{enumerate*}[label=\textbf{(\arabic*)}]
	\item We formally study MS attacks for medical image classification models under a realistic threat model of 5000 queries and hard-label access.
	\item We propose a novel MS method for training a thief model leveraging both labeled and unlabeled data. We effectively utilize supervision from an anchor model trained exclusively on labeled data, and a teacher model to make use of the thief model's unlabeled data. This ensures our technique successfully mounts MS attacks with low query budgets and without access to data from the victim model's distribution.
	\item Our evaluation of SOTA model stealing defenses reveals their failure to consistently defend against different types of MS attacks, underscoring the pressing need to address the serious implications of MS attacks in medical imaging. 
\end{enumerate*}


\section{Threat Model}

\myfirstpara{Hard labels}
We investigate attacks on victim models $f_\textrm{V}$ trained on data distribution $P_V(X)$ for medical image classification. The output of the model $y \in \{1, \dots K\}$ is a distribution over $K$ classes. The attacker has black-box query access to $f_\textrm{V}$, allowing it to submit an image $x$ and receive the models's prediction. Unlike many MS attacks, which assume access to the full probability vector $f_V(x) \in \mathbb{R}^K$, we consider a stricter scenario where only the topmost prediction $\argmax_{i\in\{1, \dots, K\}} f_V(x)_i$ from the victim is available. Additionally, the attacker lacks knowledge of the victim model's hyperparameters. 

\mypara{Different thief and victim data distributions}
Most MLaaS medical imaging models are trained on confidential patient data inaccessible to adversaries. Consequently, thief models must be trained on proxy datasets. While typical MS attacks \cite{orekondy2019knockoff,pal2020activethief} use large-scale public datasets of natural images like ImageNet \cite{russakovsky2015imagenet}, these datasets are ill-suited for our task due to their dissimilarity to medical images. Instead, malicious actors can exploit freely available medical imaging datasets online for stealing models. Therefore, a more natural choice for attackers is to use a dataset $P_A(X)$ of publicly accessible unlabeled images from the same modality. The attacker selects a subset of images from $P_A(X)$ which it queries from the victim model, thus constructing a labeled set $\mathcal{D}_{l}=\{(x_{l}^{(i)}, y_{l}^{(i)})\}_{i=1}^{N_{l}}$ of size $N_{l}$ which is then used to train the thief model $f_{T}$.

\begin{figure*}[t]
	\centering
	\includegraphics[width=0.95\linewidth]{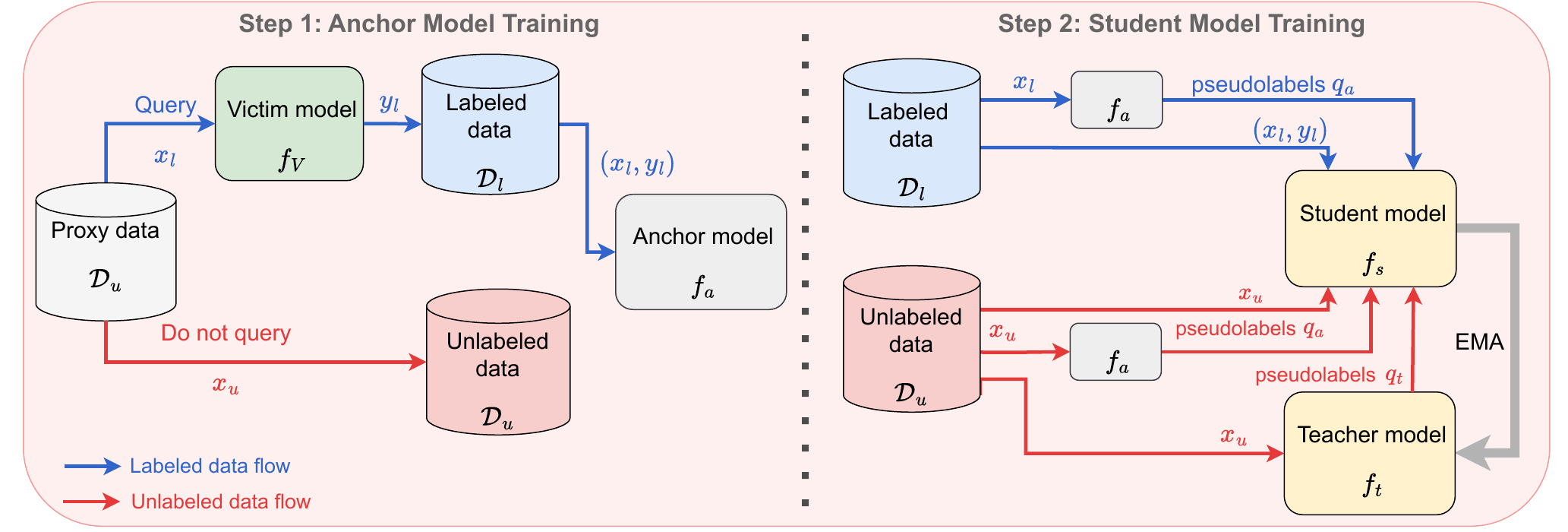}
	\caption{\textbf{Overview of \querywise:} First, a subset of the proxy data is queried from the victim model, and an \textit{anchor} model is trained on the labeled data. Our main contribution lies in the second step, where a \textit{student} model is trained using both labeled and unlabeled data. On labeled data, the student receives supervision from hard labels obtained from the victim and soft labels generated by the {anchor} model. On unlabeled data, pseudolabels are generated using both the anchor model and a \emph{teacher} model (which is updated at the end of each iteration from the student model using exponential moving average). The student model serves as the final thief model.}
	\label{fig:ssl_proposed}
\end{figure*}

\section{Proposed Method}

\Cref{fig:ssl_proposed} shows an overview of the proposed model stealing method. In addition to the conventional MS process of querying the proxy data and training the thief model on the labeled set, we additionally train on the unlabeled data as well. The proposed stealing process is carried out in two steps, as discussed below.

\subsection{Query set selection and Anchor model training} 

In the first step, we train a fully supervised thief model using the limited labeled data $\mathcal{D}_{l}$, and call it the anchor model, $f_a$. Note that the thief only needs to curate unlabeled images; these are labeled later by querying the victim model. Existing MS attacks use concepts like active learning \cite{pal2020activethief} and reinforcement learning \cite{orekondy2019knockoff} to iteratively select the query set and train the thief model. We re-purpose existing techniques to train our anchor model. 

\subsection{Student model training} \label{student_training}

The main contribution of our method lies in the second step, which is the training of student model. Our strategy is inspired by Knowledge Distillation \cite{hinton2015distilling}, but extends it to incorporate training using labeled as well as unlabeled data. Knowledge from labeled data is embedded in the fixed anchor model trained in the first step, while an additional teacher model, described below, incorporates knowledge from labeled and unlabeled data both, and is dynamically updated with the student. This additional training on the unlabeled data using artificially generated pseudo-labels (by anchor and teacher models) helps the student model surpass the anchor's performance, as can be verified from \Cref{tab1} in our experiments. Given a training mini-batch comprising $B_{l}$ labeled samples and $B_{u}$ unlabeled samples, an aggregate loss $\L = \L_{l} + \lambda \L_{u}$ is computed by combining labeled and unlabeled losses, as discussed below.

\mypara{Labeled loss, $\boldsymbol{\L_l}$} 
%
The loss is computed from the labeled samples in a mini-batch. It is a weighted combination of cross-entropy loss $\L_\text{CE}$ computed from the hard labels queried from the victim, and knowledge distillation loss $\L_\text{KD}$ computed from the anchor model's output probability vectors on the labeled data. This extra supervision from the anchor model provides regularization benefits similar to those provided by Self-Distillation \cite{furlanello2018born}.  
\begin{align} 
	\mathcal{L}_{l} &= (1-\alpha)\mathcal{L}_{\mathrm{CE}} + \alpha \mathcal{L}_{\text{KD}}, \qquad \text{where} 
	\label{eq:labeled_loss}
	\\
	\mathcal{L}_{\text{CE}} &= \sum_{i=1}^{B_{l}} \mathcal{H} \big( \sigma(q^i_{s}), y^i_{l} \big), \qquad \text{and}
	\label{eq:labeled_ce}
	\\
	\mathcal{L}_{\text{KD}} &=  \sum_{i=1}^{B_{l}} \tau^2 ~ \text{KL} \! \left( \sigma \left( \frac{q^i_{s}}{\tau} \right), \sigma \left( \frac{q^i_{a}}{\tau} \right) \right). 
	\label{eq:labeled_kd}
\end{align}
Here $y^i_{l}$ are the hard labels queried from the victim model, $q^i_s = f_s(x_{l}^i)$ and $q^i_a = f_a(x_{l}^i)$ denote the logits produced by the student and anchor models respectively for the labeled data, and $\sigma(\cdot)$ is the softmax function. $\mathcal{H}$ is the cross-entropy loss and $\text{KL}(\cdot, \cdot)$ is KL-divergence loss, with distillation temperature \cite{hinton2015distilling} $\tau>1$ used to smoothen the anchor's predictions. $\alpha \in (0,1]$ controls the relative importance of the two losses. 

\mypara{Unlabeled loss, $\boldsymbol{\L_u}$} 
The unlabeled loss tries to match the student's predictions with the softened softmax outputs of both the teacher model and the anchor model via a weighted sum of two knowledge-distillation loss components: $\L_\text{KD}^t$ computed using the teacher model, and $\L_\text{KD}^a$ using the anchor model.
\begingroup
\allowdisplaybreaks
\begin{align}\label{eq:unlabeled_loss}
	\L_u &= (1-\beta) \L_\text{KD}^t + \beta \L_\text{KD}^a, \qquad \text{where} 
	\\
	\L_\text{KD}^t &=  \sum_{i=1}^{B_u} \mathds{1} \left( \sigma(q^i_{a}) > \rho \right) 
	\tau^2 ~ \text{KL} \!
	\left( 
	\sigma \left(\frac{q^i_{s}}{\tau} \right), 
	\sigma \left( \frac{q^i_{t}}{\tau} \right) 
	\right), \qquad \text{and}  
	\label{eq:unlabeled_kdt}
	\\
	\L_\text{KD}^a &=  \sum_{i=1}^{B_{u}} \mathds{1} \left( \sigma(q^i_{a}) > \rho \right) \tau^2 ~ \text{KL} \!
	\left( 
	\sigma \left(\frac{q^i_{s}}{\tau} \right), 
	\sigma \left(\frac{q^i_{a}}{\tau} \right) 
	\right).
	\label{eq:unlabeled_kda}
\end{align}
\endgroup
Here $q^i_s = f_s(x_{u}^i)$, $q^i_t = f_t(x_{u}^i)$, $q^i_a = f_a(x_{u}^i)$ denote the logits produced by the student, teacher and anchor models respectively for the unlabeled data. 
The relative contribution of the two KD loss components from the teacher and anchor is controlled by the hyper-parameter $\beta \in (0,1]$. To curb the effect of noisy pseudo-labels \cite{sohn2020fixmatch}, only samples with the maximum softmax probability above a threshold $\rho$ are allowed to contribute to the unlabeled loss.

\mypara{Role of teacher model} 
The teacher model is initialized from the weights of the anchor model, and after each mini-batch, it is updated using an exponential moving average of the student's weights \cite{tarvainen2017mean}, defined by $	\theta_t \leftarrow m \theta_t + (1-m) \theta_s$
where $\theta_s$ and $\theta_t$ are the student's and teacher's weights respectively, and $m$ is the smoothing factor. The motivation for introducing the teacher model is to prevent the student from confirmation bias on a particular batch. At the same time, the teacher model also escapes rigidity of the anchor model by slowly updating itself from the student model based on the training on unlabeled data. The anchor and teacher models thus complement each other, and represent a balance between exploitation and cautious exploration in our method. Recall that at the end of the training procedure, thief model is derived from the student model.

\mypara{Handling class imbalance} 
We observe that the labeled set obtained by querying the victim model on an out-of-distribution thief dataset is often class-imbalanced. This may result in the thief model generalizing poorly on the less-frequent classes. 
To encourage equal contribution from all classes, we incorporate Logit Adjustment (LA) \cite{menon2020long} by applying label-dependent offsets to the student's logits $q^i_s$ during computation of the labeled loss ${\L_l}$ while training the student. 

\section{Experiments and Results}

\myfirstpara{Model stealing setup}
We use two victim models to demonstrate our model stealing capabilities. 
\begin{enumerate*}[label=\textbf{(\arabic*)}]
	\item The first is \textbf{RadFormer} \cite{basu2023radformer}, a transformer-based model for classification of \textbf{Gallbladder Cancer} (GBC) from Ultrasound (US) images. It is trained on the Gallbladder Cancer Ultrasound (GBSU) dataset \cite{basu2022surpassing} comprising of 1255 US images from 218 patients, categorized into 432 normal, 558 benign, and 265 malignant images. We use the publicly available model shared by the authors as our victim model. For stealing, we use the publicly available GBC US videos dataset \cite{basu2022unsupervised}, consisting of 32 malignant and 32 non-malignant videos containing 12,251 and 3,549 frames respectively, as our proxy dataset. Note that we do not use the video labels, but instead query the frames from the victim model and use the acquired labels instead. 
	\item The second victim model is a ResNet18 trained on \textbf{POCUS} dataset \cite{born2021accelerating} for \textbf{COVID-19 classification} that consists of 2116 images, of which 655, 349, and and 1112 are of COVID-19, bacterial pneumonia, and healthy control respectively. 
	A subset of the public COVIDx-US dataset \cite{COVIDxUS2021} comprising 13032 lung US images is used as the proxy dataset.
	For both tasks, we use a query budget of 5000 samples. Note that a budget of 5000 is much lower in comparison to existing MS attacks on general vision tasks \cite{pal2020activethief,Beetham2023} that require millions of queries. Further, model stealing requires only unlabeled images for querying, which are easier to curate than labeled images, even in medical settings.
	
\end{enumerate*}

\mypara{Training details} 
We train our thief models using a query budget of $5000$ samples, of which $10\%$ samples are set aside as validation data for hyper-parameter selection, and the rest are used for training. \querywise training incorporates two stages. 
\textbf{First stage:} The anchor model in the first stage can be obtained from any of the baseline MS methods. In our experiments, we obtain the anchor model using two baseline methods: KnockoffNets' Random selection \cite{orekondy2019knockoff} and ActiveThief's k-Center method \cite{pal2020activethief}. k-Center training is done for 5 cycles, with the training budget split uniformly across cycles. Following \cite{pal2020activethief}, we evaluate the F1 score on the validation set at the end of each epoch, and use these scores to select the best model for each k-Center training cycle. 
\textbf{Second stage:} The student model in the second stage is initialized from ImageNet-pretrained weights and trained using the same labeled set (of size 5000, queried from the victim in first stage), and train-val split as used by the anchor. We train the student model for 100 epochs ({where one epoch is defined as one pass over the labeled data, while the remaining unlabeled data is spilled over to the next epoch}).
We use $\alpha=0.4$, $\beta=0.5$, $T=1.5$,  $\lambda=1$, $\rho=0.95$, and $m=0.999$. Other training hyperparameters are listed in Supplementary Table S4.

\begin{table}[t	]
	\centering
	\caption{Model stealing performance for GBC malignancy classification task. We report accuracy (Acc.), specificity, sensitivity and agreement (Agr.). Query budget is 5000. In a significant achievement, we note that the thief model is able to outperform the radiologist accuracy using the proposed MS attack.}
	\label{tab1}
	\begin{tabular*}{\linewidth}{@{\extracolsep{\fill}}l|l|llll}
		\hline
		\textbf{Arch} & \textbf{Method} & \textbf{Acc.} & \textbf{Spec.} & \textbf{Sens.} & \textbf{Agr.}\\
		\hline
		Custom \cite{basu2022surpassing} & Victim  				& 90.16 & 90.00 & 92.86 & - \\
        \hline
        & Radiologist A \cite{basu2022surpassing} &  70.00 & 87.30 & 70.70 & - \\
        & Radiologist B \cite{basu2022surpassing} & 68.30 & 81.10 & 73.20 & - \\
		\hline
		\multirow{6}{*}{ResNet50 \cite{he2016deep}} 
		& Random \cite{orekondy2019knockoff} 	& 66.39 & 85.00 & 61.90 & 71.31 \\
		& k-Center \cite{pal2020activethief} 	& 71.31 & 87.50 & 71.43 & 68.85 \\
		& Random+FixMatch \cite{sohn2020fixmatch} & 65.57 & 82.00 & 62.00 & 66.39 \\
		& Random+\qw	& 71.31 & 80.00 & \textbf{81.00} & 74.59 \\
		& k-Center+QW	& \textbf{{72.95}} & 79.00 & \textbf{81.00} & 80.33 \\
		\hline
		\multirow{2}{*}{Inception-v3 \cite{szegedy2016rethinking}} 
		& Random \cite{orekondy2019knockoff} & 69.67 & 71.25 & \textbf{83.33} & 71.31 \\
		& Random+QW & \textbf{70.49} & 74.00 & 74.00 & 72.13 \\
		\hline
		\multirow{2}{*}{ViT \cite{dosovitskiy2020image}}
		& Random \cite{orekondy2019knockoff}			& 62.30 & 81.25 & 66.67 & 67.21 \\
		& Random+QW	 & \textbf{65.57} & 76.00 & \textbf{69.00} & 72.13 \\
		\hline
		\multirow{2}{*}{DeiT \cite{touvron2021training}}
		& Random \cite{orekondy2019knockoff}			& 71.31 & 81.25 & 78.57 & 74.59 \\
		& Random+QW		& \textbf{77.05} & 76.00 & \textbf{90.00} & 77.05 \\
		\hline
	\end{tabular*}
\end{table}

\mypara{Comparison with Baselines} We compare \querywise (\qw) to existing MS techniques: Knockoff Nets' Random selection \cite{orekondy2019knockoff}, and k-Center from ActiveThief \cite{pal2020activethief}.
We report model stealing performance for the GBC malignancy classification task in \Cref{tab1}. In addition to the standard evaluation metrics of accuracy, specificity and sensitivity, we report \textit{agreement}, which measures how often the thief's prediction matches the victim's. 
For ResNet50 thief architecture, we implement \querywise alongside two anchor models obtained from Random and k-Center MS attacks respectively. It can be seen that proposed method outperforms the respective anchor models, as well as a popular semi-supervised learning technique FixMatch \cite{sohn2020fixmatch}, in terms of both accuracy and sensitivity. We also evaluate the impact of varying the thief architecture, and observe that transformer-based networks like DeiT \cite{touvron2021training} increases model stealing accuracy as well as sensitivity. Given a fixed architecture, the proposed method (QW+Random) outperforms the respective anchor model (Random) in terms of both accuracy and agreement. \textbf{In a siginificant achievement, we show that a thief model using the proposed model stealing attack is able to outperform Radiologists' accuracy (70\% against 77\% using our MS attack), indicating the serious threat posed by the proposed attack on the proprietary models.}
We report additional results on the model stealing performance for the COVID-19 classification task in supplementary Table S1, and for natural image classifiers (trained on datasets like CIFAR-10 and Caltech-256) in supplementary Table S2, demonstrating the generality of our method across victim models.

\begin{figure}[t]
	\includegraphics[width=\linewidth]{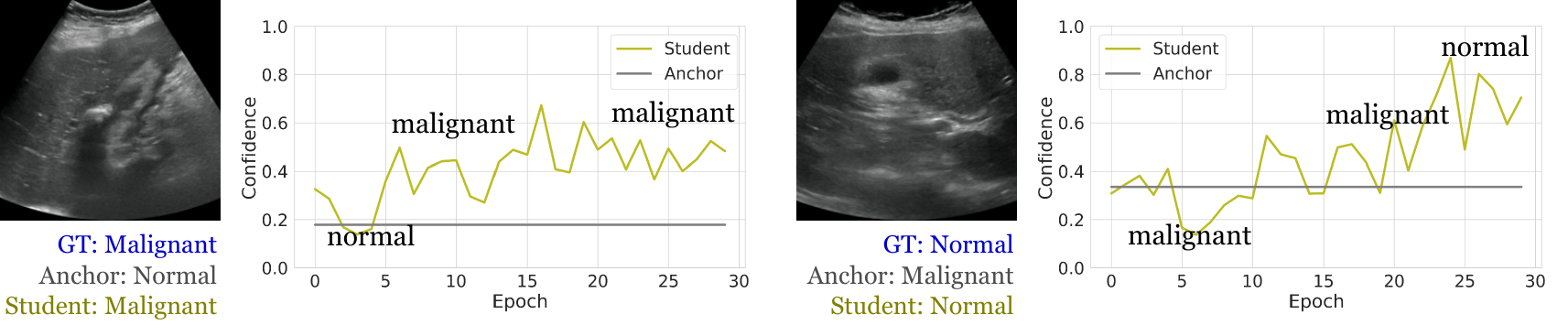}
	\caption {Unlabeled data improves thief accuracy: Comparing predictions of an \textit{anchor} model with a \textit{student} model on two GBC test set images, we show the evolution of student model's confidence for the ground truth (GT) class over the course of training, along with predicted labels. Anchor model's confidence is also shown for reference. Initially aligned with the anchor, the student gains confidence in the GT class with the use of unlabeled data, eventually making correct predictions.  } 
	\label{fig:pseudolabels}
\end{figure}

\mypara{Qualitative Comparison}
\Cref{fig:pseudolabels}  
shows the role of unlabeled data in improving thief model accuracy. We compare the predictions of anchor model with the student model, and show instances where the latter overcomes the mistakes made by the former model by learning from unlabeled data. 

\mypara{Effectiveness of Defenses}
We evaluate MS attacks on medical imaging models against SOTA MS defense techniques, in particular perturbation-based defenses that modify the victim model's output vector to prevent an attacker from accurately copying the model. These defenses trade-off victim model's accuracy for security by perturbing the predictions of the model. Three defense techniques are evaluated: Maximum Angular Deviation (MAD) \cite{orekondy2019prediction} ($\epsilon=0.8$), Adaptive Misinformation (AM) \cite{kariyappa2020defending} with detection threshold $\tau=0.7$ and Gradient Redirection (GRAD$^2$) \cite{mazeika2022steer} with $\epsilon=0.8$. 
We observe that none of the evaluated defenses consistently reduce thief accuracy for all MS methods. Moreover, any drop in thief accuracy is usually accompanied by a drop in victim accuracy, thus questioning the utility of these defenses. 
Detailed results are in suppl. Table S3.


\section{Conclusion}
We investigated the susceptibility of deployed deep-learning medical imaging models to model stealing attacks within realistic constraints, including limited query budgets and hard-label access. We proposed a new query-efficient method that effectively utilizes unlabeled data from the thief's proxy dataset alongside labeled data queries to enhance thief model performance under low query budgets. Our method proves superior to existing model stealing baselines at stealing medical imaging models. Our research invalidates the common belief regarding the safety of MLaaS framework to prevent against theft of proprietary information, and highlights the need for robust defenses against model stealing attacks.

\begin{credits}
	\subsubsection{\ackname} We acknowledge and thank the funding support from AIIMS Delhi-IIT Delhi Center of Excellence in AI funded by Ministry of Education, government of India, Central Project Management Unit, IIT Jammu with sanction number IITJMU/CPMU-AI/2024/0002. We would also like to thank Prof Vikram Goyal, Akshit Jindal, and Rakshita Choudhary for their valuable inputs to the research. 
	
	\subsubsection{\discintname}
	The authors have no competing interests to declare that are
	relevant to the content of this article. 
\end{credits}

\clearpage
\bibliographystyle{splncs04}
\bibliography{msa_ref_medical}

\clearpage
\setcounter{page}{1}
\setcounter{table}{0}
\renewcommand{\thetable}{S\arabic{table}}%
\setcounter{figure}{0}
\renewcommand{\thefigure}{S\arabic{figure}}%
\begin{center} 
{\Large Supplementary Material}
\end{center}

\begin{table}[h]
	\centering
	\caption{Model stealing performance for COVID-19 classification task. We report total accuracy (Total), class-wise accuracies for all 3 classes, and agreement (Agr.). Query budget is 5000. Proposed method achieves thief accuracy close to the baselines, while having the best agreement value.}
	\label{tab2}
	\begin{tabular*}{\linewidth}{@{\extracolsep{\fill}} l|l|lllll}
		\hline
		\textbf{Arch} & \textbf{Method} & \textbf{Total} & \textbf{COVID-19} & \textbf{Pneumonia} & \textbf{Regular} & \textbf{Agr.}\\
		\hline
		Victim & - 			 & 89.91 & 83.43 & 95.40 & 92.24 & - \\
		\hline
		\multirow{3}{*}{ResNet-50\cite{he2016deep}} 
		& Random \cite{orekondy2019knockoff} & 65.97 & 40.13 & 74.71 & 80.17 & 70.59  \\
		& k-Center \cite{pal2020activethief} & 65.55 & 57.32 & 68.97 & 69.83 & 68.49 \\
		& Random+QW	 & 63.87 & 33.12 & 72.41 & 81.47 & 71.22 \\
		\hline
	\end{tabular*}
\end{table}

\begin{table*}[h]
	\caption{Model extraction performance on general vision tasks for natural images with 5000 queries. The proposed method is implemented with two different anchor models: Random+\qw and k-Center+\qw. The best method for each dataset is depicted in \textbf{bold}, and the next best is \underline{underlined}. 
	Proposed method outperforms the baselines in terms of both accuracy and agreement for all datasets. 
Note: Dual Students\cite{Beetham2023} is a data-free method that uses synthetically generated data instead of a proxy dataset, but requires millions of queries. We implement \cite{Beetham2023} with a budget of 500K queries for the smaller datasets, yet it fails to match the performance of the other methods operating at 5000 queries. } 
	\label{tab:extraction}
	\resizebox{\linewidth}{!}{%
		\centering
		\begin{tabular}{lc|cccccccccccc}
			\hline
			\multirow{2}{*}{Method} & \multirow{2}{*}{Venue} &\multicolumn{2}{c}{\textbf{MNIST}} & \multicolumn{2}{c}{\textbf{SVHN}} & \multicolumn{2}{c}{\textbf{CIFAR10}} & \multicolumn{2}{c}{\textbf{Caltech256}} & \multicolumn{2}{c}{\textbf{CUBS200}} & \multicolumn{2}{c}{\textbf{Indoor67}}  \\
			& & Acc & Agr & Acc & Agr & Acc & Agr & Acc & Agr & Acc & Agr & Acc & Agr \\
			\hline
			Random\cite{orekondy2019knockoff} 		& CVPR'19	  & \underline{80.55} & \underline{80.59} & 67.54 & 67.84 & 65.89 & 66.66 & 39.77 & 40.32 & 14.74 & 15.75 & 33.36 & 36.22 \\
            Entropy\cite{pal2020activethief} 	& AAAI'20      & 80.55 & 80.59 & 41.02 & 41.11 & 47.53 & 48.16 & 38.88 & 39.78 & 13.87 & 15.06 & 35.82 & 39.33 \\

            k-Center\cite{pal2020activethief} 	& AAAI'20      & 71.92 & 71.99 & \underline{72.78} & \underline{73.25} & {68.02} & {68.71} & {44.66} & {45.16} & \underline{18.57} & \underline{20.14} & \underline{40.37} & \underline{42.91} \\
			
            BBD~+~Random \cite{wang2021black}  					& ECCV'22    & 27.49 & 27.51 & {58.87} & {59.04} & 47.04 & 47.59 & 40.56 & 41.16 & 16.00 & 16.64 & 34.40 & 38.06 \\
			
            BBD~+~k-Center \cite{wang2021black} 				& ECCV'22	& 58.53 & 58.52 & 43.99 & 44.18 & 40.59 & 40.58 & 41.72 & 42.05 & 16.48 & 17.41 & 30.07 & 33.88 \\
			
            Dual Students \cite{Beetham2023} 					& ICLR'23	& 18.62 & 19.24 & 6.69 & 10.89 & 12.86 & 10.16 & - & - & - & - & - & -\\
			\hline
			
            Random~+~\qw 							&	& {80.00} & {80.08} & {70.83} & {71.20} & \underline{73.07} & \underline{73.62} & \underline{45.19} & \underline{45.36} & 14.67 & 15.43 & 36.12 & 38.63 \\
			
            k-Center~+~\qw				& & \textbf{85.08} & \textbf{86.01} & \textbf{76.58} & \textbf{76.92} & \textbf{74.85} & \textbf{74.78} & \textbf{50.48} & \textbf{50.00} & \textbf{20.21} & \textbf{21.32} & \textbf{42.24} & \textbf{43.73} \\
			\hline
		\end{tabular}
	}
\end{table*}

\begin{table}[h]
\centering
\caption{Thief model accuracy under SOTA model stealing defenses. We evaluate three MS attacks on GBC malignancy classification victim model, under three defense techniques. Note that the RadFormer victim model is non-differentiable, rendering it infeasible for the defenses to compute gradients. Hence, for this experiment, we use a differentiable version of RadFormer, containing only the global branch. As can be observed, there is no significant impact (lowering of thief accuracy) that is consistent across all MS attacks. The paper advocates more research in this topic to prevent stealing of proprietary information through this route of MS attacks.
\label{tab:defense}}
\begin{tabular*}{\linewidth}{@{\extracolsep{\fill}}l|cccc}
	\hline
	Method & No Defense & MAD \cite{orekondy2019prediction} & AM \cite{kariyappa2020defending} & GRAD$^2$ \cite{mazeika2022steer}   \\
	\hline
	Victim model							& 89.34 & 80.32 & 88.52 & 86.88 \\
	\hline
	Random 									& 75.40 & 78.68 & 62.29 & 69.67\\
	Entropy \cite{pal2020activethief} 	& 74.59 & 64.75 & 65.57 & 75.40 \\
	k-Center \cite{pal2020activethief} & 70.49 & 72.13 & 72.95 & 79.50 \\
	\hline
\end{tabular*}
\end{table}

\begin{table*}[h]
	\centering
	\caption{Training hyperparameters for anchor and student models, corresponding to the two victim models. $B_l$ and $B_u$ are mini-batch sizes for labeled and unlabeled data respectively. Input image pre-processing for ViT, DeiT and Inception-v3 includes random horizontal flip and random augmentation; for ResNet-50 includes random crop, jitter, and random horizontal flip. For student model training, we use cosine learning rate decay with warmup.
		 }
	\label{tab:hyperparams}
	\begin{tabular*}{\linewidth}{@{\extracolsep{\fill}}l|l|cccc|c}
		\hline
		& & \multicolumn{4}{c|}{GBC} & COVID-19    \\
		& & ResNet50 & Inception-v3 & ViT & DeiT & ResNet50 \\
		\hline
		\multirow{5}{*}{Anchor training} 
		& learning rate & 0.005	& 0.005 & 0.005 & 0.005 & 0.01 	\\
		& momentum 		& 0.9	& 0.9 & 0.9 & 0.9 & 0.9	\\
		& epochs		& 100	& 100	& 100 & 100 & 100\\
		& batch size	& 16	& 16	& 16 & 16 & 128 \\
		& weight decay 	& 0.0005 & 0.0005 & 0.0005 & 0.0005 & 0.0005 \\
		\hline
		\multirow{7}{*}{Student training}
		& learning rate & 0.02	& 0.01 & 0.01 & 0.01 & 0.02 \\
		& momentum 		& 0.9	& 0.9 & 0.9 & 0.9 & 0.9	\\
		& weight decay 	& 0.0005 & 0.0005 & 0.0005 & 0.005 & 0.005 \\
		& epochs       	& 100	& 100  & 100 & 100 & 100 	\\
		& warmup epochs	& 10	& 10 & 10 & 10 &10	\\
		& $B_l$        	& 16	& 16    & 16 	& 16 & 16	\\
		& $B_u$       	& 112	& 112  	& 48	& 48 & 112 \\
		\hline
	\end{tabular*}
\end{table*}

\end{document}